\documentclass[prd,amsmath,superscriptaddress,nofootinbib,showpacs,twocolumn,preprintnumbers]{revtex4}
\usepackage{bm}
\usepackage{amsmath}
\usepackage{graphicx}
\usepackage{color}
\begin{document}
\title{
A microscopic study of pion condensation within Nambu--Jona-Lasinio model}

\author{R. Anglani}\email{anglani@anl.gov}
\affiliation{Physics Division, Argonne National Laboratory, Argonne, IL 60439, USA}
\affiliation{Istituto Nazionale di Fisica Nucleare (INFN), Sezione di
Bari, I-70126 Bari, Italy}
\affiliation{Dipartimento di Fisica, Universit\`a di Bari, I-70126 Bari,
Italy}

\preprint{BA-TH/624-10}
\preprint{PHY-12545-TH-2009}

\begin{abstract}
We have studied the phenomenology of pion condensation in 2-flavor neutral quark matter at finite density with Nambu--Jona-Lasinio (NJL) model of QCD. We have discussed the role of the bare quark mass $m$ and the electric chemical potential $\mu_{e}$ in controlling the condensation. The central result of this work is that the onset for $\pi$-condensed phase occurs when $|\mu_{e}|$ reaches the value of the in-medium pion mass $M_{\pi}$ provided the transition is of the second order, even for a composite pion system in the medium. Finally, we have shown that the condensation is extremely fragile with respect to the explicit chiral symmetry breaking via a finite current quark mass.
\end{abstract}

\maketitle 

\section{Introduction}
The possibility of pion condensation in nuclear medium was suggested by A.~B.~Migdal for the first time in the 1970s \cite{Migdal,picond}.
Thereafter, many studies \cite{Kunihiro:1993pz} of ``in-medium'' pion properties have been performed due to the important consequences in subnuclear physics \cite{Kienle:2004hq}, in the
 physics of neutron stars \cite{Maxwell:1977zz}, supernovas
 \cite{Ishizuka:2008gr},
 and the heavy ion collisions \cite{Zimanyi:1979ga}.
These analysis considered the pion as an elementary object but we know
that they can be considered as Nambu-Goldstone bosons generated by chiral symmetry breaking. Hence the internal structure and the mass of the pion can be sensitive to the Quantum Chromodynamics (QCD) vacuum modifications in the finite density environment \cite{Hatsuda:1994pi}; moreover the finite baryon density, and even the isospin density arising from the neutrality condition, can modify the structure of the QCD ground state, and  it can in turn produce significant modifications of the pion properties in the medium.

In this work~\cite{Abuki:2008wm},
 we present the possibility of a charged pion
($\pi^c$) condensation starting from a {\em microscopic} model which is built with quarks as the constituents of pions, and which exhibits chiral symmetry restoration at the finite quark chemical potential $\mu$ or temperature $T$.
We derive an appropriate criterion for $\pi^c$ condensation, that we use to show the fragility of the $\pi^c$ condensation with respect to explicit chiral symmetry breaking via a finite current quark mass. The conditions for the onset of $\pi^c$ condensation at finite density are investigated using the Nambu--Jona-Lasinio (NJL) model of QCD. We find that the threshold for $\pi^c$ condensation for noninteracting elementary pion gas is $\mu_e = M_{\pi^-}$ ($-\mu_e=M_{\pi^+}$) for positive (negative) $\mu_e$, where $\mu_e$ is the electric chemical potential and $M_{\pi}$ the in-medium pion mass. 
Finally, we clarify the effect of the current quark mass in the electrically neutral ground state, and the effect of a non-vanishing $\mu_e$ in presence of a finite current quark mass. Based on these analyses, we conclude that $\pi^c$ condensation is forbidden in realistic neutral quark matter within this effective model.

\section{The model}
The NJL Lagrangian of the model is given for a two-flavor quark matter at a finite chemical potential~\cite{Abuki:2008tx}
\begin{eqnarray}
{\cal L} &=& \bar{e}(i\gamma_\mu\partial^\mu + \mu_e \gamma_0)e +
 \bar\psi\left(i\gamma_\mu \partial^\mu + \hat\mu\gamma_0 -m\right)\psi
 \nonumber \\
&&+ G\left[\left(\bar\psi \psi\right)^2 + \left(\bar\psi i \gamma_5
        \vec\tau \psi\right)^2\right]~, \label{eq:Lagr}
\end{eqnarray}
Here $e$ denotes the electron field, and $\psi$ is the quark spinor with
Dirac, color and flavor indices (implicitly summed). The bare quark mass is $m=m_u=m_d$ and $G$ the coupling constant. The electrical chemical potential $\mu_e$ is necessary to keep the system
 electrically neutral \cite{Abuki:2008tx}, while $\mu_I$ serves as the
 isospin chemical potential in the hadron sector, $\mu_I=-\mu_e$.
The quark chemical potential matrix $\hat\mu$ is defined in flavor-color
 space as $\hat\mu=\text{diag}(\mu-\frac{2}{3}\mu_e,
 \mu+\frac{1}{3}\mu_e)\otimes\bm{1}_c$,
 where $\bm{1}_c$ denotes the identity matrix in color space, $\mu$ is
 the quark chemical potential related to the conserved baryon number.
In the mean field approximation (MFA), we examine the possibility
 that the ground state develops condensation in the $\sigma =
 G\langle\bar\psi\psi\rangle$
 and/or ${\bm \pi}=G\langle\bar\psi i\gamma_5{\bm \tau}\psi\rangle$
 channels, where ${\bm \tau}=\{\tau_1,\tau_2,\tau_3\}$ denotes the Pauli
 matrices. 
Due to the absence of driving forces, we find $\langle\pi_3\rangle$ is always zero.
For convenience we denote $M=m-2\sigma$ and $N=2\sqrt{\pi_1^2+\pi_2^2}$.
In the numerical analyses, we have fixed $\Lambda=651$ MeV and $G=2.12/\Lambda^2$ so that the model reproduces $f_\pi=92$ MeV, $\langle\bar{u}{u}\rangle=-(250\,{\rm MeV})^3$, and $m_\pi=139$ MeV in the vacuum with $m=5.5$ MeV.

\section{Pion condensation at finite $\mu_I$ and at finite $\mu$}
The pion condensation ``in the vacuum'' ($\mu=0$) has been investigated with the chiral perturbation approach~\cite{Son:2000by} and within NJL model~\cite{He:2005nk}. In both studies the threshold for the onset of the condensation is found to be $\mu_{e}=m_{\pi}$, i.e., when the absolute value of the electric chemical potential equals the vacuum pion mass.

\begin{figure}[t]
\begin{center}
\includegraphics[width=7cm]{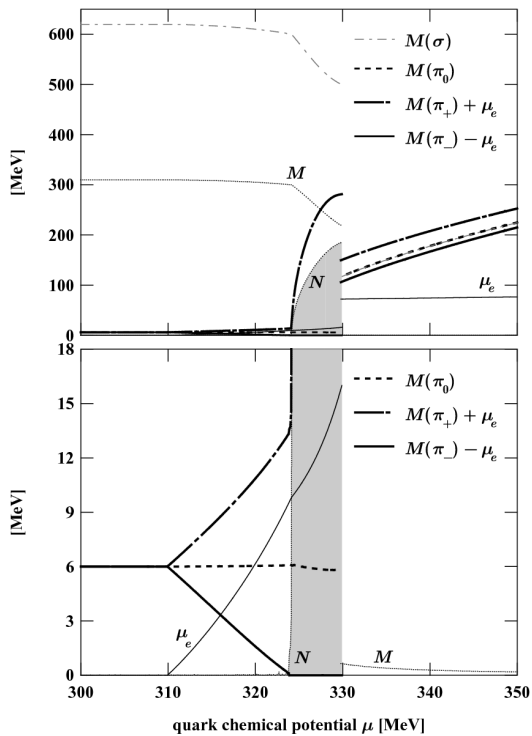}
\end{center}
\caption{
The constituent quark mass $M$, the pion
 condensate $N$, and meson masses as a function of $\mu$
 at $T=0$ in the neutral phase for a toy value of the current 
 quark mass $m=10$ keV.}
\label{fig:masse}
\end{figure}

We now consider neutral quark matter at $\mu\neq0$ and $T=0$ and we want to study the relation between the threshold of $\pi^c$
 condensation at finite density and the in-medium pion masses in the
 neutral ground state (for further recent studies at finite baryon and/or isospin chemical potential see also Refs.~\cite{Ebert:2009ty,Matsuzaki:2009kb}).
In Ref.~\cite{Abuki:2008tx}, it is shown that at the physical
 point $m=5.5$ MeV, there is no room for $\pi^c$ condensation in the
 neutral phase (similar results obtained also in Ref.~\cite{Andersen:2007qv}). Even though the picture changes when the current mass is lowered, for our discussion it is enough to state that we consider a current
 quark mass of the order of 10 keV. In Fig.~\ref{fig:masse} we plot $M$ and $N$ in the neutral phase as a function of $\mu$.
In this figure $M_{\pi^0}$, $M_{\pi^\pm}$ denote the in-medium pion
 masses defined by the poles of the pion propagators in the rest frame, computed in the randomt phase approximation (RPA) to the Bethe-Salpeter (BS) equation.
The positive and negative solutions of the BS equation in $\omega$ corresponds to the excitation gaps for $\pi^+$ and $\pi^-$, which are  $(M_{\pi^+}+\mu_e)$ and $(M_{\pi^{-}}-\mu_e)$, respectively. From Fig.~\ref{fig:masse} we notice that the transition to the pion
 condensed phase is of second order and it occurs at the point where
 $M_{\pi^-} = \mu_e$. For a more detailed mathematical discussion of the numerical results shown in Fig.~\ref{fig:masse} we refer to~\cite{Abuki:2008wm}.

\section{The role of the current mass and the phase diagram $(\mu,\mu_{e})$}
In the past years, a large number of the analysis about $\pi^c$ condensation have been performed in the chiral limit.
In this section we investigate the role of the finite current quark mass in $\pi^c$ condensation. In order to do this, we set the cutoff $\Lambda$ and the coupling $G$ to the values specified above and we treat $m$ as a free parameter. As a consequence, the pion mass at $\mu=T=0$, $m_\pi$ in the following, is a free parameter as well.
\begin{figure}[t]
\begin{center}
\includegraphics[width=7cm]{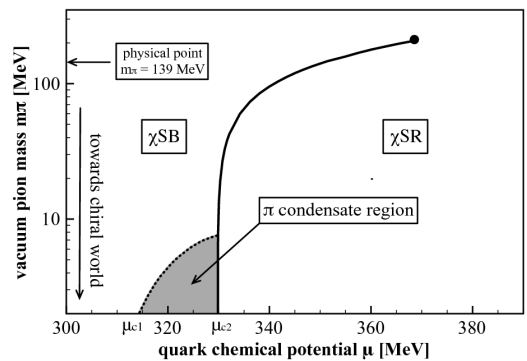}
\end{center}
\caption{Phase diagram of neutral matter in $(\mu,\,m_{\pi})$ plane.}
\label{fig:phasempi}
\end{figure}
In Fig.~\ref{fig:phasempi} we report the phase diagram in
 $(\mu,\,m_\pi)$ plane in the neutral case.
The solid line represents the border between the two regions where 
chiral symmetry is broken and restored.
The bold dot is the critical endpoint of the first order transition.
The shaded region indicates the region where $\pi^c$ condensation
 occurs.
In the chiral limit $(m_{\pi}=0)$ our results are in good agreement with
 those obtained in Ref.~\cite{Ebert:2005wr}.
Indeed, there exist two
 critical values of the quark chemical potential, $\mu_{c1}$
 and $\mu_{c2}$, corresponding to the onset and vanishing of
 $\pi^c$ condensation, respectively.
When the current quark mass increases, a shrinking of the shaded
 region occurs till the point $\mu_{c1}\equiv \mu_{c2}$
 for $m_{\pi}^{c}\sim 9\; \rm MeV$, corresponding to a current quark
 mass of $m \sim 10\;\rm keV$.
Hence the gapless $\pi^c$ condensation is
 extremely fragile with respect to the symmetry breaking effect of 
 the current quark mass.
\begin{figure}[t]
\begin{center}
\includegraphics[width=7cm]{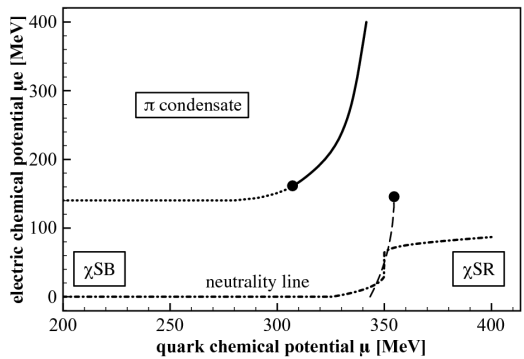}
\end{center}
\caption{
Phase diagram in $(\mu,\,\mu_{e})$ plane
 at $m=5.5$ MeV.}
\label{fig:phasediag} 
\end{figure}
As a final investigation, in Fig.~\ref{fig:phasediag} we report the phase
 diagram of quark matter in the $(\mu,\,\mu_e)$ plane when the current quark mass is tuned to $m=5.5$ MeV.
At each value of $(\mu,\,\mu_e)$ we compute the chiral and pion
 condensates by minimization of the thermodynamical potential.
The solid line represents the first order transition from the $\pi^c$
 condensed phase to the chiral symmetry broken phase without the $\pi^c$
 condensate.
The bold dot is the critical endpoint for the first order transition,
 after which the second order transition sets in.
The dashed line indicates the first order transition between the two
 regions where chiral symmetry is broken and restored, respectively.
The dot-dashed line is the neutrality line
 $\mu_e^{\rm neut}=\mu_e(\mu)$
 which is obtained by requiring the global electrical neutrality condition,
 $\partial\Omega/\partial\mu_e = 0$.
The neutrality line manifestly shows the impossibility of finding a $\pi^c$ condensate, in this physical situation.

\section{Conclusions} 
We have studied the pion condensation in two-flavor neutral quark matter using the Nambu--Jona-Lasinio model of QCD at finite density. We have investigated the role of electric charge neutrality,
and explicit symmetry breaking via quark mass, both of which control the onset of the charged pion condensation. We show that the equality between the electric chemical potential and the
``in-medium'' pion mass, $|\mu_{e}|=M_{\pi} $, as a threshold, persists even for a composite pion system in the
medium, provided the transition to the pion condensed phase is of the second order (the same qualitative picture has been found also in a recent analysis in the massive Gross-Neveu model \cite{Ebert:2009ty}). Moreover, we
have found that the pion condensate in neutral quark matter is extremely fragile to the symmetry breaking
effect via a finite current quark mass $m$, and is ruled out for $m$ larger than the order of 10 keV.

A final comment is in order. In this study the contribution to finite baryon density comes only from the constituent quarks and not from nucleons. The presence of nucleons may make the pion condensation even less favorable. The reason can be found in the Tomozawa-Weinberg pion-nucleon interaction, which is isospin odd, giving rise to a repulsive pion self-energy at the physical situation where the neutron density is larger than the proton density (see Ref.~\cite{Muto:2003dk}). For this reason an interesting investigation may consider the pion condensation with the presence of nucleon degrees of freedom within NJL model (see Ref.~\cite{Bentz:2002um} for useful prescriptions).

This work was supported in part by the U.S. Department of Energy, Office of Nuclear Physics, under contract no. DE-AC02-06CH11357.

\end{document}